\documentclass[preprint,showpacs,preprintnumbers,amsmath,amssymb]{revtex4}

\usepackage{graphicx}
\usepackage{dcolumn}
\usepackage{bm}

\begin{document}

\title{Measurement of Pancharatnam's phase by robust interferometric and polarimetric
methods}
\author{J. C. Loredo$^{1}$, O. Ort\'{i}z$^{1}$, R. Weing\"{a}rtner$^{1,2}$ and F. De
Zela$^{1}$}

\affiliation{$^{1}$ Departamento de Ciencias, Secci\'{o}n F\'{i}sica \\
Pontificia Universidad Cat\'{o}lica del Per\'{u}, Apartado 1761, Lima, Peru.%
\\
$^{2}$ Department of Materials Science 6,\\
University of Erlangen-N\"{u}rnberg, Martensstr. 7, 91058
Erlangen, Germany}

\begin{abstract}
We report theoretical calculations and experimental observations
of Pancharatnam's phase originating from arbitrary $SU(2)$
transformations applied to polarization states of light. We have
implemented polarimetric and interferometric methods which allow
us to cover the full Poincar\'{e} sphere. As a distinctive
feature, our interferometric array is robust against mechanical
and thermal disturbances, showing that the polarimetric method is
not inherently superior to the interferometric one, as previously
assumed. Our strategy effectively amounts to feed an
interferometer with two copropagating beams that are orthogonally
polarized with respect to each other. It can be applied to
different types of standard arrays, like a Michelson, a Sagnac, or
a Mach-Zehnder interferometer. We exhibit the versatility of our
arrangement by performing measurements of Pancharatnam's phases
and fringe visibilities that closely fit the theoretical
predictions. Our approach can be easily extended to deal with
mixed states and to study decoherence effects.

\end{abstract}


\maketitle

\section{\label{sec:level1}Introduction}

As is well known, Pancharatnam's phase was originally introduced
to deal with the relative phase of two polarized light beams
\cite{pancharatnam}. It anticipated geometrical phases that are
nowadays intensively studied both theoretically and
experimentally. Among all geometrical phases, Berry's phase
\cite{berry} has played a major role in prompting the upsurge of a
vast amount of investigations dealing with topological phases in
quantum and classical physics. Berry's phase was originally
introduced by considering the adiabatic evolution of a quantum
state subjected to the action of a parameter-dependent
Hamiltonian. However, the first experiments aiming at exhibiting
such a phase were performed with classical states of light, using
cw-lasers \cite{tomita}. It was soon realized that the phase
tested in such experiments differed from Berry's phase, as it was
larger than the latter by a factor of two. The reason for this was
that the experimentally studied phase \cite{tomita} arose from
$SO(3)$ instead of $SU(2)$ transformations. Indeed, Tomita and
Chiao \cite{tomita} let polarized light pass a coiled optical
fiber and measured the phase originated from the adiabatic change
suffered by the propagation direction of a light-beam. Thus, the
corresponding parameter space being explored -- the sphere of
directions -- differed from the parameter space that was involved
in Berry's original phase. The latter was Bloch sphere, on which
any spin-$1/2$ state can be represented. Another two-state system
formally equivalent to a spin-$1/2$ state is polarized light, in
which, e.g., vertically ($V$) and horizontally ($H$) polarized
states constitute the counterparts of the spin-up and spin-down
quantum states. Polarization states can be represented on the
Poincar\'{e} sphere, which is equivalent to the Bloch sphere. An
early experiment testing the appearance of Pancharatnam's phase in
polarization states describing closed paths on the Poincar\'{e}
sphere was the one performed by Bhandari and Samuel
\cite{bhandari}. This interferometric test was however restricted
to a limited set of $SU(2)$ transformations and, moreover, some of
the transformations used by the authors were nonunitary, as they
employed linear polarizers to bring the polarization back to its
initial value. Thus, Chyba \emph{et al.} \cite{chyba} performed
alternative tests by employing only unitary transformations to
exhibit Pancharatnam's phase, though such transformations were
still restricted to cover a limited $SU(2)$ range. In spite of its
original formulation in terms of polarization states of light,
Pancharatnam's phase has not been fully exhibited in optical
implementations, in contrast to more recent experiments based on
neutron spin-interferometry \cite{wagh3,wagh4,wagh5}. Some years
ago, Wagh and Rakhecha \cite{wagh1,wagh2} proposed two alternative
methods to measure Pancharatnam's phase. One method is based on a
polarimetric procedure, while the other is an interferometric one.
Both procedures have been tested and
compared against one another in experiments using neutrons \cite{wagh3,wagh4}%
. The conclusion drawn from these experiments was that the
polarimetric method is inherently superior to the interferometric
method. This is so mainly because the polarimetric method is
insensitive to mechanical and thermal disturbances that usually
plague interferometric methods. Neutron interferometry, in
particular, is also limited through spacial constraints that are
imposed by the geometry of the monocrystals used to construct the
interferometers. In order to explore a large range in the
parameter space of the geometric phase, people contrived to
realize some regions of this space by electrically inducing
phase-changes that were beyond the range accessible through
rotation of a flipper. However, such a procedure prompted some
criticisms \cite{bhandari2} concerning the parameter spaces that
were involved in the two phase evolutions, as one of them was
physically obtained by rotation of a flipper and the other by
electrical means. On the other hand, the allegedly more accurate
polarimetric method allows phase measurements only modulo $\pi $,
and is therefore unable to verify certain features like the
anticommutation of Pauli matrices, e.g., $\sigma _{x}\sigma
_{y}=-\sigma _{y}\sigma _{x}$, something that was beautifully done
with the interferometric method \cite{wagh3}.

To the best of our knowledge, the two methods referred to above
have not yet been tested against each other in all-optical
experiments being capable of exploring the full parameter range of
the Poincar\'{e} sphere. We have thus endeavor to compare both
methods of measuring Pancharatnam's phase by using all-optical
setups. In this work we present a robust interferometric
arrangement that makes accessible the full range of $SU(2)$
polarization transformations. Furthermore, we have also
implemented a polarimetric array with a similar coverage, so that
both methods could be compared against each other. As we shall
see, our interferometric arrangement is insensitive to mechanical
and thermal disturbances. This represents an important
improvement, as compared to conventional interferometric
arrangements. The latter are usually set up as a variant of a
Michelson, a Mach-Zehnder, or a Sagnac interferometer. Our method
works with any of these variants, so that one could choose the
most appropriate arrangement. For example, one could explore
decoherence effects by measuring geometric phases in polarization
single-photon mixed-states using Mach-Zehnder interferometers,
similarly to recently reported experiments \cite{ericsson}. In
such a case, the fringe contrast (visibility) of the
interferometric pattern also conveys information about the
geometric phase. Though our work deals with pure states only, we
have also tested the visibility of our patterns as a function of
$SU(2)$ transformations, obtaining very good agreement with
theoretical predictions.

Our experiments, in addition to test Pancharatnam's phase with great
versatility, serve the purpose of showing a common ground for classical and
quantum manifestations of topological phases. Indeed, although our tests
have been performed with classical states of light, they could be
straightforwardly extended to experiments with single photons. Our
theoretical discussion has thus been couched in a quantum-mechanical
language, so that, e.g., the polarization states of classical light are
represented by kets like $|V\rangle $ and $|H\rangle $. It should thus be
clear that the features under study are not of an intrinsic classical or
quantum-mechanical nature. Instead, it is the topological aspect what
manifests itself as a common ground for both classical and quantum phenomena.

The paper is organized as follows. In Section \ref{sec:level2} we
review the interferometric and the polarimetric methods for
measuring Pancharatnam's phase and derive theoretical results that
apply in our case. In Section \ref{sec:level5} we describe our
experimental arrangements and present our results, comparing them
with our theoretical predictions. Finally, we present in Section
\ref{sec:level8} our conclusions.

\section{\label{sec:level2}The interferometric and polarimetric methods}

Given two states, $\left| i\right\rangle $ and $\left| f\right\rangle $,
their Pancharatnam's relative phase $\Phi _{P}$\ is defined as $\Phi
_{P}=\arg \left\langle i|f\right\rangle $. A very direct way to exhibit $%
\Phi _{P}$ is through interferometry. Indeed, consider two interfering,
non-orthogonal states, $\left| i\right\rangle $ and $\left| f\right\rangle $%
, with $\left| i\right\rangle \neq \left| f\right\rangle $. If we apply a
phase shift $\phi $ to one of the states, the resulting intensity pattern is
given by

\begin{equation}
I=\left| e^{i\phi }\left| i\right\rangle +\left| f\right\rangle
\right| ^{2}=2+2\left| \left\langle i|f\right\rangle \right| \cos
\left( \phi -\arg \left\langle i|f\right\rangle \right) .
\label{i1}
\end{equation}
The maxima of $I$ are thus attained at $\phi =\arg \left\langle
i|f\right\rangle =\Phi _{P}$. We are interested in exhibiting
$\Phi _{P}$ in two-level systems and when Pancharatnam's phase
arises as a consequence of having submitted an initial state
$\left| i\right\rangle $ to an arbitrary transformation $U\in
SU(2)$ that converts it into a final state $\left| f\right\rangle
=U\left| i\right\rangle $. The intensity measurement for which Eq.
(\ref{i1}) applies can be implemented with the help of, say, a
Mach-Zehnder interferometer. Alternatively, one could employ
polarimetric methods. We will discuss both methods in what
follows. But before, and for later reference, let us introduce the
two parametrizations of $U\in SU(2)$ that we shall use in our
analysis. We call them the $YZY$-form and the $ZYZ$-form, for
obvious reasons: The first one is given by
%



\begin{equation}
U(\xi ,\eta ,\zeta )=\exp \left( -i\frac{\xi }{2}\sigma _{y}\right) \exp
\left( i\frac{\eta }{2}\sigma _{z}\right) \exp \left( -i\frac{\zeta }{2}%
\sigma _{y}\right) ,  \label{u2}
\end{equation}
while the second form is given by
\begin{equation}
U(\beta ,\gamma ,\delta )=\exp \left( i(\frac{\delta +\gamma }{2})\sigma
_{z}\right) \exp \left( -i\beta \sigma _{y}\right) \exp \left( i(\frac{%
\delta -\gamma }{2})\sigma _{z}\right) =\left(
\begin{tabular}{ll}
$e^{i\delta }\cos \beta $ & $-$ $e^{i\gamma }\sin \beta $ \\
$e^{-i\gamma }\sin \beta $ & $e^{-i\delta }\cos \beta $%
\end{tabular}
\right) .  \label{u3}
\end{equation}

To pass from one form of $U$ to the other one needs to connect the
respective parameters. The corresponding equations of
transformation involve, generally, trigonometric formulae, so that
the different parameters are not connected to one another through
algebraic relations. The representation of Eq. (\ref{u3}) is
particularly well adapted to exhibit Pancharatnam's phase. Indeed,
taking as initial state $\left| i\right\rangle =\left|
+\right\rangle _{z}$, i.e., the eigenstate of $\sigma _{z}$ that
belongs to the eigenvalue $+1$, and setting $\left| f\right\rangle
=U\left| +\right\rangle _{z}$ we have

\begin{equation}
\left\langle i|f\right\rangle = \ _{z} \!\left \langle +\right|
U(\beta ,\gamma ,\delta )\left| +\right\rangle _{z}=e^{i\delta
}\cos \beta .  \label{d}
\end{equation}

From the definition of Pancharatnam's phase, i.e., $\Phi _{P}=\arg
\left\langle
i|f\right\rangle $, we obtain $\Phi _{P}=\delta +\arg (\cos \beta )$, for $%
\beta \neq (2n+1)\pi /2$. Because $\cos \beta $ can take on positive and
negative real values, $\arg (\cos \beta )$ equals $0$ or $\pi $, and $\Phi
_{P}$ is thus defined modulo $\pi $. In any case, the parametrization $%
U(\beta ,\gamma ,\delta )$ of Eq. (\ref{u3}) is seen to be most
appropriate to exhibit $\Phi _{P}=\delta $ (modulo $\pi $). On the
other hand, for the optical implementation of $U$ the
parametrization of the $YZY$-form is more appropriate. Indeed, it
is well known \cite{simon0} that with the help of three retarders,
viz, two quarter-wave plates and one half-wave plate, it is
possible to implement any $U\in SU(2)$ in the polarization-space
of, e.g., horizontally and vertically polarized states of light:
$\left\{ \left| H\right\rangle ,\left| V\right\rangle \right\} $.
This requires that one represents $U$ in the form given by Eq.
(\ref{u2}), i.e., the $YZY$-form, because of the following
relationship involving the Euler angles $\theta _{1}$, $\theta _{2}$, $%
\theta _{3}$ (see, e.g., \cite{englert}):

\begin{equation}
\exp \left( -i(\theta _{3}+3\pi /4)\sigma _{y}\right) \exp \left(
i(\theta _{1}-2\theta _{2}+\theta _{3})\sigma _{z}\right) \exp
\left( i(\theta _{1}-\pi /4)\sigma _{y}\right) =Q(\theta
_{3})H(\theta _{2})Q(\theta _{1}). \label{p1}
\end{equation}
Here, $Q$ means a quarter-wave plate and $H$ a half-wave plate.
The arguments of the retarders are the angles of their major axes
to the vertical direction. In the case of a $U$ given by Eq.
(\ref{u2}), by applying Eq. (\ref{p1}) we obtain

\begin{equation}
U(\xi ,\eta ,\zeta )=Q\left( \frac{-3\pi +2\xi }{4}\right) H\left( \frac{\xi
-\eta -\zeta -\pi }{4}\right) Q\left( \frac{\pi -2\zeta }{4}\right) .
\label{p2}
\end{equation}

Having discussed the two parametrizations of $U\in SU(2)$ that are
useful for our purposes, we turn now to the implementation of the
experimental arrangements that allow us to exhibit Pancharatnam's
phase.

\subsection{\label{sec:level3}Interferometric arrangement: Mach-Zehnder and Sagnac}

In general, with an interferometric array Pancharatnam's phase can
be drawn from intensity measurements that are essentially
described by Eq. (\ref{i1}). If we introduce $U$ as given in Eq.
(\ref{u2}) into Eq. (\ref{i1}), we obtain the intensity as

\begin{eqnarray}
I &=&\left| \frac{1}{\sqrt{2}}\left( e^{i\phi }\left| +\right\rangle
_{z}+U(\xi ,\eta ,\zeta )\left| +\right\rangle _{z}\right) \right| ^{2}=
\label{i2} \\
&=&1-\cos \left( \frac{\eta }{2}\right) \cos \left( \frac{\xi +\zeta }{2}%
\right) \cos \left( \phi \right) -\sin \left( \frac{\eta }{2}\right) \cos
\left( \frac{\xi -\zeta }{2}\right) \sin \left( \phi \right) .  \notag
\end{eqnarray}

From Eqs. (\ref{u2}) and (\ref{u3}) it follows that the parameters
of these two representations of $U\in SU(2)$ are related through
$\tan (\delta )=\tan \left( \frac{\eta }{2}\right) \cos \left( \frac{%
\xi -\zeta }{2}\right) / \cos \left( \frac{\xi +\zeta
}{2}\right)$. Hence, $I$ can be written as

\begin{equation}
I=1-\cos \left( \frac{\eta }{2}\right) \cos \left( \frac{\xi +\zeta }{2}%
\right) \sec \left( \delta \right) \cos \left( \delta -\phi \right) ,
\label{i3}
\end{equation}
making it evident that an interferometric method for exhibiting
$\Phi _{P}$ would require measuring the shift induced by $U$ on
the intensity pattern by an angle $\delta =\Phi _{P}$ (modulo $\pi
$). Now, the expression for $I$ as given in Eq. (\ref{i3}) is
somewhat inconvenient, because it mixes $\delta $ with parameters
of a representation to which it does not belong. By expressing Eq.
(\ref{i3}) with the parameters of $U(\beta ,\gamma ,\delta )$ we
obtain

\begin{equation}
I=1-\cos \left( \beta \right) \cos \left( \delta -\phi \right) ,
\label{i3a}
\end{equation}
thus rendering clear that the visibility $v\equiv \left( I_{\max
}-I_{\min }\right) /\left( I_{\max }+I_{\min }\right) $ is given
by $v=\cos \beta $, i.e., it is independent of Pancharatnam's
phase. In terms of the parameters $\xi $, $\eta $, $\zeta $ the
square of the visibility is given by

\begin{equation}
v^{2}(\xi,\eta,\zeta) =\frac{1}{2}\left[ 1+\cos \xi \cos \zeta -\cos \eta
\sin \xi \sin \zeta \right] .  \label{vis1}
\end{equation}

For experimental tests, it will be useful to write the visibility in terms
of the angles of the retarders:

\begin{eqnarray}
&&v^{2}(\theta _{1},\theta _{2},\theta _{3})  \label{vis2} \\
&=&\frac{1}{2}\left[ 1+\cos \left( \frac{3\pi +4\theta _{3}}{2}\right) \cos
\left( \frac{\pi -4\theta _{1}}{2}\right) -\cos \left( 2\theta _{1}-4\theta
_{2}+2\theta _{3}\right) \sin \left( \frac{3\pi +4\theta _{3}}{2}\right)
\sin \left( \frac{\pi -4\theta _{1}}{2}\right) \right] .  \notag
\end{eqnarray}

Let us now refer specifically to a Mach-Zehnder interferometer. In
order to calculate its output intensity, let us represent light
beams as a superposition of polarization states ($\left\{ \left|
H\right\rangle ,\left| V\right\rangle \right\} $) and momentum (or
``which way'', i.e., spatial) states ($\left\{ \left|
X\right\rangle ,\left| Y\right\rangle \right\} $). These last
states denote the two-way alternative that can be ascribed to the
Mach-Zehnder interferometer. Let us consider first that our
initial state is taken to be a vertically polarized state that
enters the first beam-splitter along the $X$-direction (e.g., the
beam passing polarizer $P_{1}$ in Fig. (1)). It is represented by
$\left| VX\right\rangle \equiv \left| V\right\rangle \otimes
\left| X\right\rangle $. The actions of beam-splitters, mirrors
and phase-shifters are represented by operators in the two-qubit
space with basis $\{\left| VX\right\rangle ,\left| VY\right\rangle
,\left| HX\right\rangle ,\left| HY\right\rangle \}$. They act on
the $\left| X\right\rangle ,\left| Y\right\rangle $ states,
leaving the polarization states $\left| H\right\rangle ,\left|
V\right\rangle $ unchanged. The action of a $50:50$ beam splitter
and the action of a mirror are given, respectively, by
\cite{englert}
\begin{eqnarray}
U_{BS} &=&1_{P}\otimes \frac{1}{\sqrt{2}}\left( \left| X\right\rangle
\left\langle X\right| +\left| Y\right\rangle \left\langle Y\right| +i\left|
X\right\rangle \left\langle Y\right| +i\left| Y\right\rangle \left\langle
X\right| \right) , \\
U_{mirr} &=&1_{P}\otimes \left[ -i\left( \left| X\right\rangle \left\langle
Y\right| +\left| Y\right\rangle \left\langle X\right| \right) \right] ,
\end{eqnarray}
with $1_{P}$ meaning the identity operator in polarization space.
Let us stress that the above expressions for the actions of a
beam-splitter and a mirror hold true irrespective of the fact that
the spatial qubits be realized by classical or by quantum fields
(see, e.g., \cite{schleich}). Working with classical fields, the
usage of kets (and bras) is just a useful mathematical means to
represent field amplitudes. Accordingly, a phase factor in one or
in the other arm of the interferometer can be represented by
$U_{X}(\phi )=1_{P}\otimes \left( \exp \left( i\phi \right) \left|
X\right\rangle \left\langle X\right| +\left| Y\right\rangle
\left\langle Y\right| \right) $ and $U_{Y}(\phi )=1_{P}\otimes
\left( \left| X\right\rangle \left\langle X\right| +\exp \left(
i\phi \right) \left| Y\right\rangle \left\langle Y\right| \right)
$, respectively. If we mount an array of retarders on, say, arm
$X$ of the interferometer, its action would be represented by
$U_{P}^{X}= U\otimes \left| X\right\rangle \left\langle
X\right| +1_{P}\otimes \left| Y\right\rangle \left\langle Y\right| $%
, where $U\in SU(2)$ means the respective polarization
transformation that the retarders produce, in our case the one
given in Eq. (\ref{u2}). Similarly, $U_{P}^{Y}=1_{P}\otimes \left|
X\right\rangle \left\langle X\right|+ U\otimes \left|
Y\right\rangle \left\langle Y\right| $. For the arrangement shown
in Fig. (1) we obtain

\begin{equation}
U_{T}=U_{BS}U_{mirr}U_{X}(\phi )U_{P}^{Y}U_{BS}.  \label{t}
\end{equation}

This $U$ acts on the initial state $\left| VX\right\rangle $ and
the intensity measured at one of the output ports of the final
beam-splitter is obtained by projecting the resulting state,
$U_{T}\left| VX\right\rangle $, with the appropriate projectors:
$\left| VX\right\rangle \left\langle VX\right| $ and $\left|
HX\right\rangle \left\langle HX\right| $, thereby obtaining the
vectors $\left| VX\right\rangle \left\langle VX\right| U_{T}\left|
VX\right\rangle $ and $\left| HX\right\rangle \left\langle
HX\right| U_{T}\left| VX\right\rangle $, respectively. Squaring
the respective amplitudes and summing up we get the intensity as
$I_{V}=\left| \left\langle VX\right| U_{T}\left| VX\right\rangle
\right| ^{2}+\left| \left\langle HX\right| U_{T}\left|
VX\right\rangle \right| ^{2}$. A straightforward calculation
yields

\begin{equation}
I_{V}=\frac{1}{2}\left[ 1-\cos \left( \frac{\eta }{2}\right) \cos \left(
\frac{\xi +\zeta }{2}\right) \cos \left( \phi \right) -\sin \left( \frac{%
\eta }{2}\right) \cos \left( \frac{\xi -\zeta }{2}\right) \sin \left( \phi
\right) \right] .
\end{equation}

As already shown, this can be written as

\begin{equation}
I_{V}=\frac{1}{2}\left[ 1-\cos \left( \beta \right) \cos \left( \phi -\delta
\right) \right] .  \label{iv}
\end{equation}

Using the above result, a direct measurement of Pancharatnam's
phase $\delta =\Phi _{P}$ (modulo $\pi $) becomes possible: all
one needs to do is to measure the fringe-shift between two
interferograms, one of them serving as reference ($\delta =0$),
and the other being obtained after applying the $U$
transformation. The practical problem with this method is the
instability of the interferometric array. Minute changes in any
component of the interferometer preclude an accurate determination
of $\delta $. Different strategies can be applied to overcome this
kind of shortcomings. A mechanical and thermal isolation of the
arrangement is the most direct one, but it makes measurements
rather awkward. Damping instabilities by a feedback mechanism is
another possibility; but it makes the arrangement more involved
and difficult to operate. A third option would be to use a Sagnac
instead of a Mach-Zehnder interferometer. In a Sagnac-like
interferometer one can make the two beams pass the same optical
elements, so that any instability would affect both beams equally.
One should then design the interferometer in such a way that the
$U$ transformation acts on one beam alone, so that the other can
serve as the reference beam. In our case, for reasons explained in
detail in Section \ref{sec:level5}, we turned to a different
option that is based on the following observations.

Eq. (\ref{iv}) holds for an initial state that is vertically
polarized. When the initial state is horizontally polarized, then
the intensity is given by

\begin{equation}
I_{H}=\frac{1}{2}\left[ 1-\cos \left( \beta \right) \cos \left( \phi +\delta
\right) \right] .  \label{ih}
\end{equation}
We observe that intensities $I_{V}$ and $I_{H}$ are shifted with respect to
each other by $2\delta $. Thus, we can exploit this fact for measuring $%
\delta $. To this end, we polarize one half -- say the upper half
-- of the laser beam vertically, and the lower half horizontally.
With such a beam we feed our interferometer, be it in a
Mach-Zehnder or in a Sagnac configuration, so that we can capture
at the output an interferogram, half of which corresponds to
$I_{V}$ and the other half to $I_{H}$. The upper
fringes of the output will be shifted with respect to the lower ones by $%
2\delta $. As both halves of the beam pass the same optical
elements, they will be equally affected by whatever perturbations.
The array is therefore insensitive to instabilities. We thus need
only to accurately measure the relative fringe-shift in each
interferogram in order to obtain $\delta $. By applying this
method we have measured Pancharatnam's phase with an accuracy that
is similar to that reached by the polarimetric method, on which we
turn next.

\subsection{\label{sec:level4}Polarimetric arrangement}

The optical setup for the polarimetric method, as proposed by Wagh
and Rakhecha \cite{wagh1}, is somewhat more demanding, as compared
to the interferometric method. At first sight, however, the
polarimetric method could appear to be the simplest of the two
options, because it requires a single beam, from which one
extracts phase information. It is not obvious that phase
information can be extracted from a single beam. However, the
polarimetric method is in fact based on an analogous principle as
the interferometric one, and in a certain sense polarimetry could
be seen as ``virtual interferometry''. Let us briefly discuss how
it works.

Consider an initial, polarized state $\left| i\right\rangle =\left|
+\right\rangle _{z}$, and submit it to the action of a $\pi /2$-rotation
around an axis perpendicular to the polarization axis ($z$), e.g., a
rotation around the $x$-axis. As a result, we obtain the state $\left(
\left| +\right\rangle _{z}-i\left| -\right\rangle _{z}\right) /\sqrt{2}$. If
we now phase-shift this state by applying to it the operator $\exp \left(
-\phi \sigma _{z}/2\right) $ we obtain the state $V\left| +\right\rangle
_{z}\equiv \exp \left( -i\phi \sigma _{z}/2\right) \exp \left( -i\pi \sigma
_{x}/4\right) \left| +\right\rangle _{z}=\left( e^{-i\phi /2}\left|
+\right\rangle _{z}-ie^{i\phi /2}\left| -\right\rangle _{z}\right) /\sqrt{2}%
=e^{-i\phi /2}\left( \left| +\right\rangle _{z}-ie^{i\phi }\left|
-\right\rangle _{z}\right) /\sqrt{2}$. We have thus generated a relative
phase $\phi $ between the states $\left| +\right\rangle _{z}$ and $\left|
-\right\rangle _{z}$, something analogous to what is achieved in an
interferometer by changing the length of one of the two optical paths.
Subsequently, we let $U\in SU(2)$ act and as a result we obtain the state $%
UV\left| +\right\rangle _{z}=\left( e^{-i\phi /2}U\left|
+\right\rangle _{z}-ie^{i\phi /2}U\left| -\right\rangle
_{z}\right) /\sqrt{2}\equiv \left| \chi _{+}\right\rangle +\left|
\chi _{-}\right\rangle $. It is from this last state that we can
extract Pancharatnam's phase by intensity measurements. In order
to accomplish this goal we project $\left| \chi _{+}\right\rangle
+\left| \chi _{-}\right\rangle $ on the state $V\left|
+\right\rangle _{z}$, i.e., the phase-shifted, split state we
prepared before applying $U$. The corresponding intensity we
measure is thus given by

\begin{equation}
I=\left| \ _{z}\left\langle +\right| V^{\dagger }\left( \left|
\chi _{+}\right\rangle +\left| \chi _{-}\right\rangle \right)
\right| ^{2}. \label{i5}
\end{equation}
Let us write $V\left| +\right\rangle _{z}=\left( e^{-i\phi /2}\left|
+\right\rangle _{z}-ie^{i\phi /2}\left| -\right\rangle _{z}\right) /\sqrt{2}%
\equiv \left| \varphi _{+}\right\rangle +\left| \varphi
_{+}\right\rangle $ and take $U$ as given by $U(\beta ,\gamma
,\delta )$ of Eq. (\ref{u3}). Calculating the amplitude
$_{z}\left\langle +\right| V^{\dagger }\left( \left| \chi
_{+}\right\rangle +\left| \chi _{-}\right\rangle \right)
=(\left\langle \varphi _{+}\right| +\left\langle \varphi
_{-}\right| )(\left| \chi _{+}\right\rangle +\left| \chi
_{-}\right\rangle )$ we obtain, using $\left\langle \varphi
_{\pm}|\chi _{\pm}\right\rangle =\exp \left(\pm i\delta \right)
\cos \left( \beta \right) /2$, and $\left\langle \varphi
_{\mp}|\chi _{\pm}\right\rangle =i\exp \left(\mp i(\gamma +\phi
)\right) \sin \left( \beta \right) /2$, that $(\left\langle
\varphi _{+}\right| +\left\langle \varphi _{-}\right| )(\left|
\chi _{+}\right\rangle +\left| \chi _{-}\right\rangle )=\cos
\left( \beta \right) \cos \left( \delta \right) +i\sin \left(
\beta \right) \cos \left( \gamma +\phi \right) $ and, hence, that
the intensity amounts to

\begin{equation}
I=\cos ^{2}\left( \beta \right) \cos ^{2}\left( \delta \right) +\sin
^{2}\left( \beta \right) \cos ^{2}\left( \gamma +\phi \right) .  \label{i6}
\end{equation}

Eq. (\ref{i6}) contains Pancharatnam's phase $\delta =\Phi _{P}$ (modulo $\pi $%
) in a form that allows its extraction through intensity
measurements. Indeed, we observe from Eq. (\ref{i6}) that the
minimal and maximal intensities are given by $I_{\min }=\cos
^{2}\left( \beta \right) \cos ^{2}\left( \delta \right) $ and
$I_{\max }=\cos ^{2}\left( \beta \right) \cos ^{2}\left( \delta
\right) +\sin ^{2}\left( \beta \right) $, respectively, so that

\begin{equation}
\cos ^{2}\left( \delta \right) =\frac{I_{\min }}{1-I_{\max }+I_{\min }},
\label{wr}
\end{equation}
which is the expression on which the polarimetric method is finally based.

A concrete experimental arrangement requires that we implement $V$ and $U$
with the help of retarders. To begin with, $\exp \left( -i\pi \sigma
_{x}/4\right) =Q(\frac{\pi }{4})$ and $\exp \left( -i\phi \sigma
_{z}/2\right) =Q(\frac{\pi }{4})H(\frac{\phi -\pi }{4})Q(\frac{\pi }{4})$.
Using $Q^{2}(\frac{\pi }{4})=H(\frac{\pi }{4})$ and $\exp \left( +i\phi
\sigma _{z}/2\right) =Q(-\frac{\pi }{4})H(\frac{\phi +\pi }{4})Q(-\frac{\pi
}{4})$ we obtain

\bigskip
\begin{equation}
U_{tot}\equiv V^{\dagger }UV=H\left( -\frac{\pi }{4}\right) H\left( \frac{%
\phi +\pi }{4}\right) Q\left( -\frac{\pi }{4}\right) UQ\left( \frac{\pi }{4}%
\right) H\left( \frac{\phi -\pi }{4}\right) H\left( \frac{\pi }{4}\right) .
\label{up}
\end{equation}
As for $U$, it is convenient to employ the form $U(\xi ,\eta
,\zeta )$ of Eq. (\ref{u2}), a form which can be directly
translated into an arrangement with retarders, according to Eq.
(\ref{p2}), i.e., an arrangement of the form $QHQ$. Inserting this
$QHQ$ for $U$ into Eq. (\ref{up}) we end up with an arrangement
that consists of nine plates. In order to reduce the number of
plates we apply relations like, e.g., $Q(\alpha )H(\beta )=H(\beta
)Q(2\beta -\alpha )$, $Q(\alpha )H(\beta )H(\gamma )=Q(\alpha +\pi
/2)H(\alpha -\beta +\gamma -\pi /2)$. The final reduction gives an
array that consists of five retarders:

\begin{eqnarray}
U_{tot} &=&Q\left( -\frac{3\pi }{4}-\frac{\phi }{2}\right) Q\left( -\frac{%
5\pi +2\xi }{4}-\frac{\phi }{2}\right) Q\left( -\frac{9\pi +2\left( \xi
+\eta \right) }{4}-\frac{\phi }{2}\right) \times  \label{ut} \\
&&\times H\left( -\frac{7\pi +\xi +\eta -\zeta }{4}-\frac{\phi }{2}\right)
Q\left( -\frac{\pi }{4}-\frac{\phi }{2}\right) .  \notag
\end{eqnarray}

Note that such an arrangement could be implemented by mounting
five plates having a common rotation axis, so that all the plates
can be rotated simultaneously by the same angle $\phi /2$. The
intensity that we should measure at the detector depends on $\xi
$, $\eta $, and $\zeta $ according to the following expression:

\begin{eqnarray}
I &=&\left| _{z}\left\langle +\right| U_{tot}\left| +\right\rangle
_{z}\right| ^{2}=  \label{i7} \\
&=&\cos ^{2}\left( \frac{\eta }{2}\right) \cos ^{2}\left( \frac{\xi +\zeta }{%
2}\right) +\left[ \cos \left( \frac{\eta }{2}\right) \sin \left( \frac{\xi
+\zeta }{2}\right) \cos \left( \phi \right) +\sin \left( \frac{\eta }{2}%
\right) \sin \left( \frac{\xi -\zeta }{2}\right) \sin \left( \phi \right) %
\right] ^{2}.  \notag
\end{eqnarray}

From this intensity we can extract Pancharatnam's phase, as given
by Eq. (\ref {wr}). We have tested this theoretical prediction
under restricted
conditions, by manually rotating the retarders. Thus, we fixed $\zeta $ to $%
2\pi $, so that $\cos ^{2}\left( \delta \right) =I_{\min }\left(
1-I_{\max }+I_{\min }\right) ^{-1}=\cos ^{2}\left( \eta /2\right) $ for all $%
\xi $. In such a case, Pancharatnam's phase (modulo $\pi $) should
be given by $\Phi _{P}=\eta /2$. For $\zeta =2\pi $ the
arrangement that realizes the corresponding $U_{tot}$ reduces to
the following expression:


\begin{equation}
U_{tot}^{\zeta =2\pi }=Q\left( \phi \right) Q\left( -\frac{\xi
}{2}+\phi \right) H\left( \frac{\eta -\xi }{4}+\phi \right) ,
\label{uz}
\end{equation}
in which we have redefined the rotation angle $\phi $ according to
$(-3\pi -2\phi )/4\rightarrow \phi $.
 If we instead fix $\xi =-\pi $, it still remains true that $\cos ^{2}\left(
\delta \right) =I_{\min }\left( 1-I_{\max }+I_{\min }\right) ^{-1}=\cos
^{2}\left( \eta /2\right) $, this time for all $\zeta $, so that $\Phi
_{P}=\eta /2$ (modulo $\pi $), as before. The corresponding arrangement of
retarders is now given by

\begin{equation}
U_{tot}^{\xi =-\pi }=Q\left( \frac{3\pi +2\eta -2\phi }{4}\right) H\left(
\frac{-4\pi +\zeta +\eta -2\phi }{4}\right) Q\left( \frac{-\pi -2\phi }{4}%
\right) .  \label{ux}
\end{equation}

 It is worth noting that the intensity in this case is given by

\bigskip
\begin{equation}
I=\cos ^{2}\left( \frac{\zeta }{2}\right) \cos ^{2}\left( \frac{\eta -2\phi
}{2}\right) +\sin ^{2}\left( \frac{\zeta }{2}\right) \cos ^{2}\left( \frac{%
\eta }{2}\right) .
\end{equation}
Setting $\eta =0$, $\zeta =\pi $ the intensity has a constant
value, which is useful for adjusting the arrangement. The results
of our measurements, including those corresponding to the full
array with five retarders, are shown in Fig. (\ref{q1}). As one
can see, they confirm the theoretical predictions within
experimental errors.

\section{\label{sec:level5}Experimental procedures and results}

\subsection{\label{sec:level6}Polarimetric measurements}

We have carried out measurements of the Pancharatnam phase by applying the
polarimetric and the interferometric methods presented in the previous
sections. In both cases the light source was a $30$ $mW$, cw He-Ne laser ($%
632.8$ $nm$). The polarimetric arrangement shown in Fig.
(\ref{f2}) could have been designed so that the five retarders
(see Eq. (\ref{ut})) could be simultaneously rotated by the same
amount. If one aims at systematically measuring Pancharatnam's
phase with the polarimetric method, this would require having a
custom-made apparatus on which one can mount the five plates with
any desired initial orientation and then submit the whole assembly
to rotation. As our aim was to simply exhibit the versatility of
the method and to compare its accuracy with that of the
interferometric method, we mounted a simple array of five
independent retarders so that each one of them could be manually
rotated. With such an approach it takes some hours of painstaking
manipulation to record all necessary data, whenever the experiment
is performed with the full array of five retarders. For this
reason, we initially restricted our tests to three retarders. This
could be achieved by lowering the degrees of freedom, i.e., by
fixing one of the three Euler angles, as explained in the previous
section (see Eqs. (\ref{uz}, \ref{ux})). Having made measurements
with three plates we performed an additional run of measurements
with the full arrangement of five retarders. Our results are shown
in Fig. (\ref{q1}). They correspond to intensity measurements
obtained with a high-sensitivity light sensor (Pasco CI-6604, Si
PIN photodiode with spectral response in the range $320$ $nm$ –-
$1100$ $nm$). As expected (retarders and polarizers could be
oriented to within $1^{0}$), the experimental values are within
$3\%$ to $6\%$ in accordance with the theoretical predictions,
depending on the number of retarders being employed.

\subsection{\label{sec:level7}Interferometric measurements}

We used two interferometric arrangements. One of them was a
Mach-Zehnder interferometer, and the other was a Sagnac
interferometer. We started by mounting both interferometers in the
standard way, but adding an array of three retarders on one arm
for implementing any desired $U\in SU(2)$. Usually, phase shifts
$\phi $, as appearing in Eq. (\ref{i3a}), originate from moving
one mirror with, e.g., a low-voltage piezotransducer. One can then
record the interference pattern by sensing the light intensity
with a photodiode set at one of the output ports of the exiting
beam-splitter. Alternatively, one can capture the whole
interference pattern with a CCD camera. The Mach-Zehnder
interferometer is easier to mount in comparison to the Sagnac
interferometer. However, it has the disadvantage of being more
unstable against environmental disturbances, thus requiring the
application of some stabilizing technique like, e.g., a feedback
system. In contrast, the Sagnac interferometer is very stable with
respect to mechanical and thermal disturbances. Nevertheless,
mounting a Sagnac interferometer can be difficult, for geometrical
reasons. By using one or the other method one can obtain two
interferograms -- one with, and the other without, the retarders
in place. In our case, capturing the whole interference pattern
with a CCD camera -- instead of sensing it with a photodiode --
proved to be the most convenient approach with both arrangements,
Mach-Zehnder and Sagnac. When working with the Mach-Zehnder array
we first implemented a feedback system in order to stabilize the
reference pattern. One of the two paths followed by the laser beam
was used for feedback. The feedback system should allow us to
compensate the jitter and thermal drifts of the fringe patterns
that preclude a proper measurement of the phase shift. This
requires that an electronic signal, after proportional-integral
amplification, be fed to a piezotransducer in a servo-loop, so as
to stabilize the interferometer, thereby locking the fringe
pattern. Although we succeeded in locking the fringe pattern, the
geometry of our array severely limited the parameter range we
could explore. We thus turned to a different option, i.e., the one
based on Eqs. (\ref{iv}) and (\ref{ih}). It required polarizing
one half of the laser beam in one direction and the other half in
a direction perpendicular to the first one.

\begin{figure}[tbp]
\begin{center}
\includegraphics[angle=0,scale=.5]{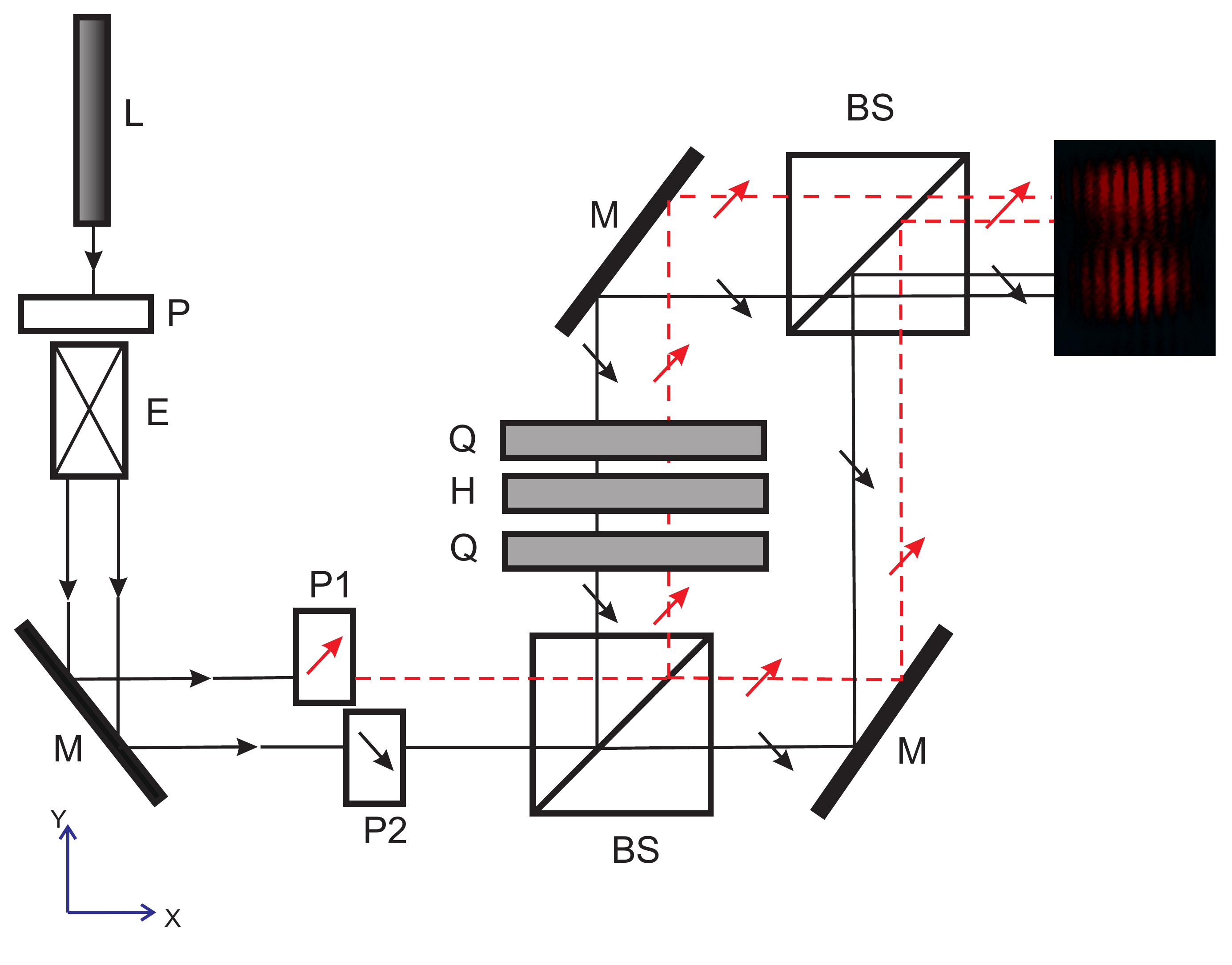}
\end{center}
\caption{(Color online) Interferometric arrangement for testing Pancharatnam's phase $%
\Phi_{P}$. Light from a He-Ne laser ($L$) passes a polarizer ($P$)
and enters a beam expander ($E$), after which half of the beam
goes through one polarizer ($P_{1}$) and the other half goes
through a second polarizer ($P_{2}$), orthogonally oriented with
respect to the first. The two collinear beams feed the same
Mach-Zehnder interferometer ($BS$: beam-splitter, $M$: mirror), in
one of whose arms an array of three retarders has been mounted
($Q$:quarter-wave plate, $H$: half-wave plate), so as to realize
any desired $SU(2)$ transformation. This transformation introduces
a Pancharatnam phase $\Phi_{P}=\delta$ on one half of the beam and
an opposite phase $\Phi_{P}=-\delta$ on the other, perpendicularly
polarized half, so that the relative phase of the two halves
equals $2\delta$. From the relative shift between the upper and
lower halves of the interferogram that is captured by a CCD camera
set at the output of the array one can determine $\Phi_{P}$. Any
instability of the array affects both halves of the interferogram
in the same way, so that the relative shift $2\delta$ is
insensitive to instabilities.} \label{f1}
\end{figure}

\begin{figure}[tbp]
\begin{center}
\includegraphics[angle=0,scale=.5]{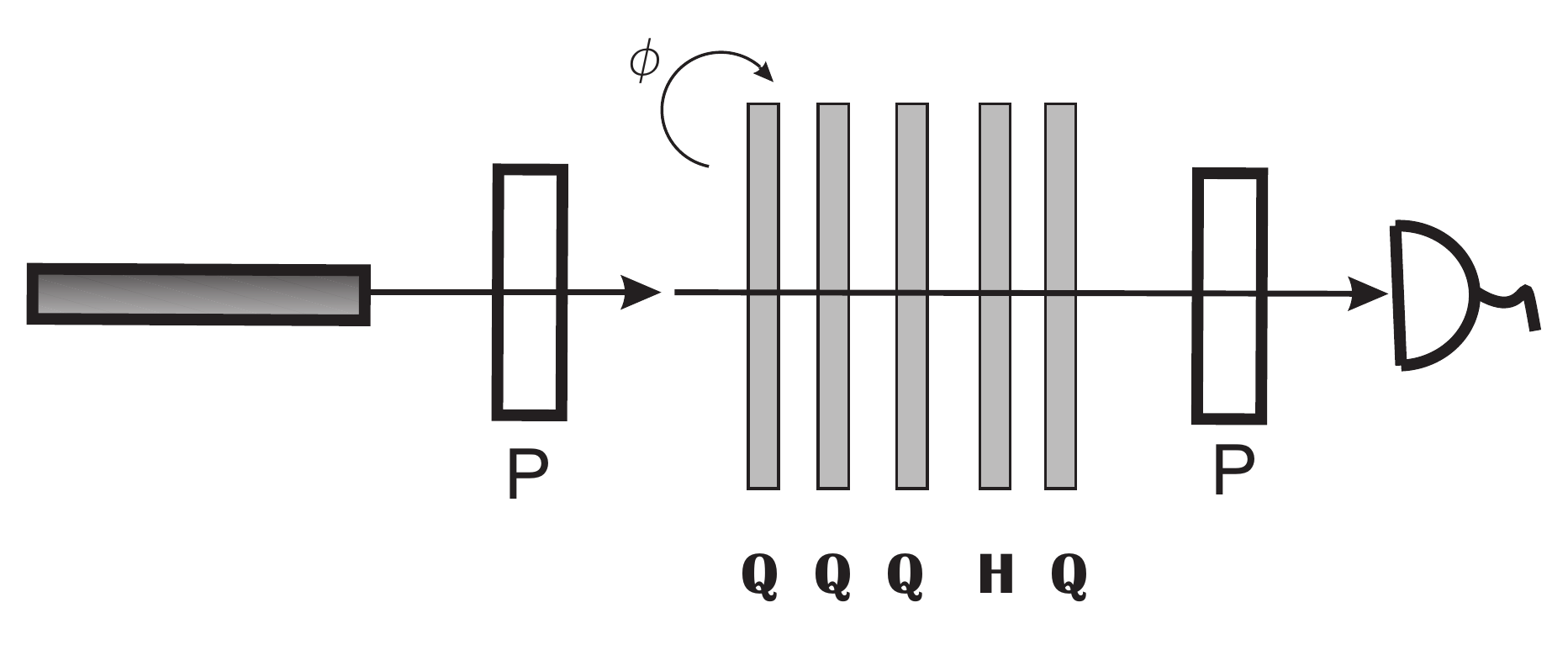}
\end{center}
\caption{Polarimetric arrangement for testing Pancharatnam's phase
 $\Phi_{P}$. With an array of five retarders ($Q$:quarter-wave plate, $H$: half-wave plate)
and two polarizers ($P$) a relative phase $\phi$ between two
polarization components $|\pm\rangle_{z}$ can be introduced, on
which any desired $SU(2)$ transformation can be applied. The five
retarders are simultaneously rotated, thereby varying $\phi$, and
the intensity $I(\phi)$ is recorded. From the maximum and minimum
values of $I$ one can determine $\Phi_{P}$, according to $\cos
^{2}\left( \Phi_{P} \right) =I_{\min }/(1-I_{\max }+I_{\min })$.}
\label{f2}
\end{figure}

\begin{figure}[tbp]
\begin{center}
\includegraphics[angle=0,scale=.6]{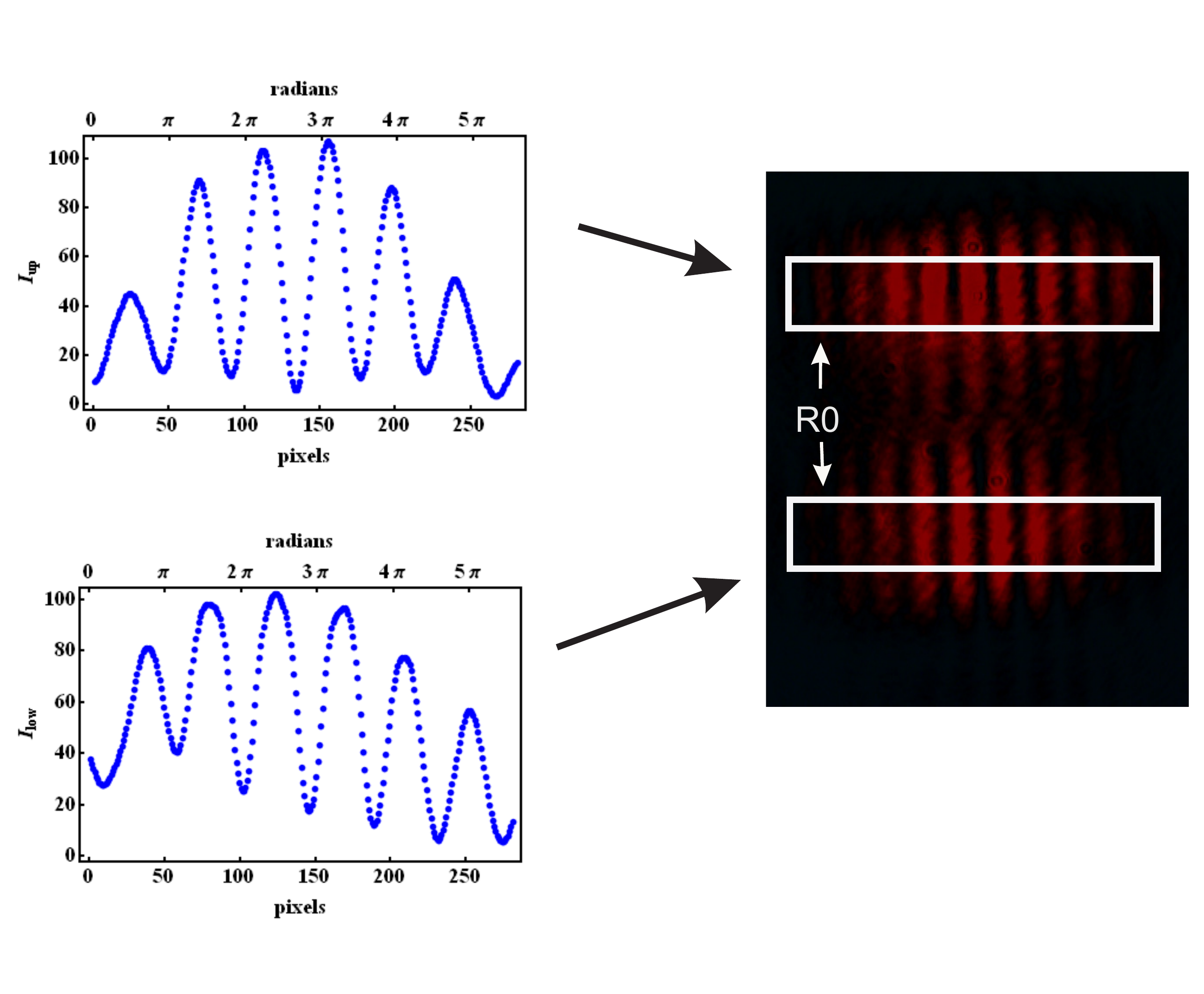}
\end{center}
\caption{(Color online) Pancharatnam's phase can be extracted from
the relative fringe-shift between the upper and lower parts of the
interferogram. The relative shift equals twice the Pancharatnam's
phase. The left panels show the result of performing a column
average of the fringes plus the application of a Savitzky-Golay
filter to get rid of noise features. The column average is
performed after selecting the evaluation area $R_{0}$ on the
interferogram, as illustrated on the right panel. The reported
shifts are mean values obtained from four different selections,
$R_{0},\ldots, R_{3}$, of the evaluation area.}
\label{interferogram}
\end{figure}

\begin{figure}[tbp]
\begin{center}
\includegraphics[angle=0,scale=.7]{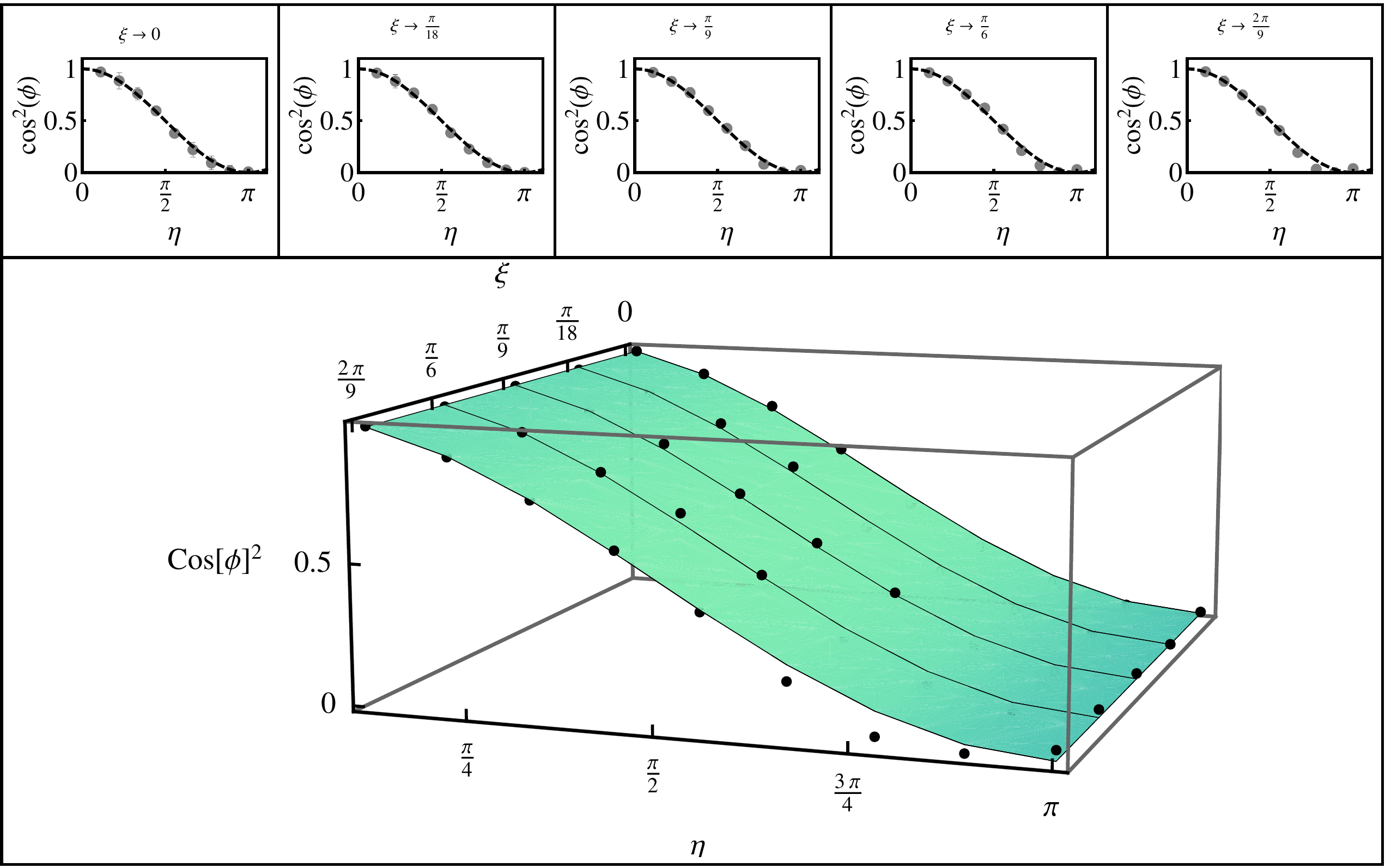}
\end{center}
\caption{(Color online) Experimental results from the
interferometric measurement of Pancharatnam's phase. We plot
$\cos^{2}(\Phi_{P})$ as a function of $\xi$ and $\eta$, with
$\zeta$ being held fixed to zero. In the upper panels we plot the
single curves that are highlighted on the surface shown on the
lower panel. Dots correspond to experimental values, some of which
fall below and some above the surface.} \label{int1}
\end{figure}

\begin{figure}[tbp]
\begin{center}
\includegraphics[angle=0,scale=1.]{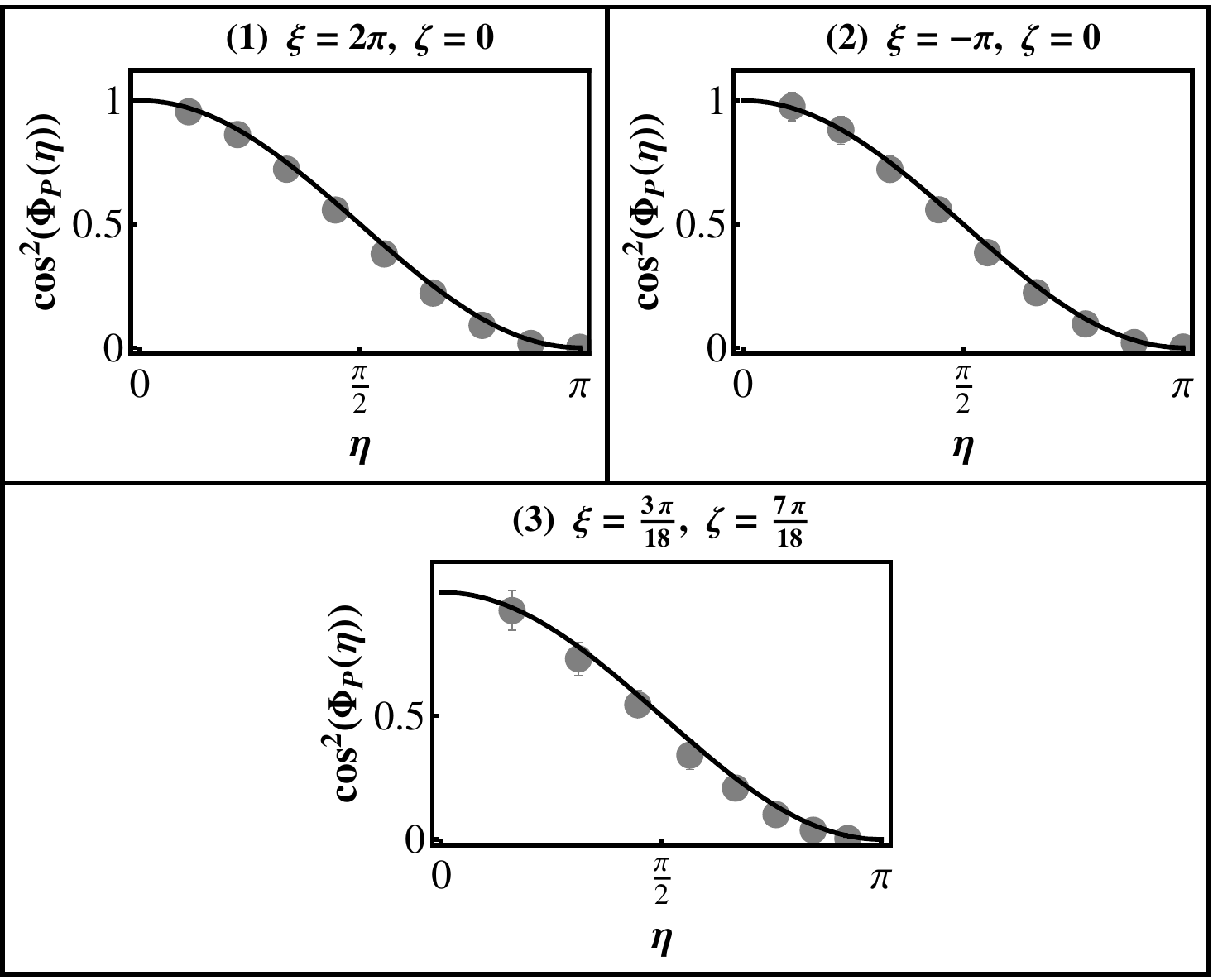}
\end{center}
\caption{Experimental results from a polarimetric measurement of
Pancharatnam's phase. The upper graphs correspond to an array that
consists of three retarders set in the form $QQH$ (left) and $QHQ$
(right). Parameter values are as indicated and
$\cos^{2}(\Phi_{P})$ was measured as a function of $\eta$. The
lower curve corresponds to the full array of five retarders set in
the form $QQQHQ$.} \label{q1}
\end{figure}

In order to exhibit the feasibility of our interferometric method
we performed experiments with both a Mach-Zehnder and a Sagnac
array. In both cases we obtained similar preliminary results.
However, the systematic recording of our results corresponds to
the Mach-Zehnder array shown in Fig. (\ref{f1}), as it was the
simpler one to mount and manipulate. As shown in the figure, the
initially polarized laser beam was expanded so that its upper half
passed through one polarizer $P_{1}$ and its lower half through a
second polarizer $P_{2}$, orthogonally oriented with respect to
the first. Each run started by setting the retarders so as to
afford the identity transformation: $Q(\pi /4)H(-\pi /4)Q(\pi
/4)=1_{P}$, the corresponding interferogram was captured with a
CCD camera ($1/4''$ Sony CCD, video format of $640\times 480$
pixels, frame rate adjusted to $30$ fps) and digitized with an
$IBM$-compatible computer. The upper and lower halves of this
interferogram showed a small relative shift stemming from surface
irregularities and tiny misalignments. The initial interferogram
served to gauge all the successive ones that correspond to
transformations $U(\xi ,\eta ,\zeta )\not=1_{P}$. Each
interferogram was evaluated with the help of an algorithm that
works as follows. First, by optical inspection of the whole set of
interferograms -- corresponding to a
given $U(\xi ,\eta ,\zeta )$ -- one selects (by pixel numbers) a common region $%
R_{0}$ of the images the algorithm should work with (see Fig.
(\ref{interferogram})). Having this region as its input the
algorithm performs a column average of each half of the
interferogram -- thereby obtaining the mean profile of the fringes
-- and the output is then submitted to a low-pass filter
(Savitzky-Golay filter) to get rid of noisy features. The result
is a pair of curves like those shown in Fig.
(\ref{interferogram}). The algorithm then searches for relative
minima in each of the two curves and compares their locations so
as to output the relative shifts between the minima of the curves.
After averaging these relative shifts the algorithm produces its
final output for each pair of curves. We repeated this procedure
for a series of regions (fixed by pixel numbers): $R_{0}\ldots
R_{3}$, so that we could estimate the uncertainty of our
experimental values. No attempt was made to automate the selection
of the working regions. Visual inspection proved to be effective
enough for our present purposes. Some series of interferograms
showed limited regions that were clearly inappropriate for being
submitted to evaluation, as they reflected inhomogeneities and
other features that stemmed from surface irregularities of the
optical components. We applied the complete procedure to a whole
set of interferograms corresponding to different choices of $U(\xi
,\eta ,\zeta )$. Our results are shown in Fig. (\ref{int1}). As
can be seen, our experimental results are in very good agreement
with theoretical predictions.

A second, independent, algorithm was also used to check the above
results. This algorithm was developed as a variant of some
commonly used procedures in image processing. Like in the previous
approach, the algorithm first constructs the mean profiles of the
fringes and submits them to a low-pass filter. But now, instead of
searching for relative minima, the algorithm does the following.
First, it determines the dominant spatial carrier frequency
$k_{0}$ by Fourier transforming curves like those shown in Fig.
(\ref{interferogram}).
Let us denote these curves by $\widehat{i}_{up}(x)$ and $\widehat{i}%
_{low}(x)$, corresponding, respectively, to the upper and lower half of the
interferogram. The Fourier transforms are denoted by $%
i_{up}(k) $ and $i_{low}(k)$. The goal is to determine the relative shift $%
\Delta _{r}=2\delta $ between $\widehat{i}_{up}(x)$ and $\widehat{i}%
_{low}(x) $. It can be shown \cite{goldberg} that $\Delta
_{r}=\Delta _{up}-\Delta _{low}\approx \Im[\log
(i_{up}(k_{0}))]-\Im[\log (i_{low}(k_{0}))]$, up to a constant
phase-offset that is the same for all the interferograms
pertaining to a given $U(\xi ,\eta ,\zeta )$. The above expression
for $\Delta _{r}$ comes from observing that both $i_{up}(k_{0})$
and $i_{low}(k_{0})$ have the structure
$i(k_{0})=a(k_{0})+b(0)\exp (i\Delta )+b^{\ast }(2k_{0})\exp
(-i\Delta )$, so that $i(k_{0})\approx b(0)\exp (i\Delta )$
whenever $\left|
b(0)\right| \gg \left| b^{\ast }(2k_{0})\right| $, $\left| a(k_{0})\right| $%
. Thus, the accuracy of the approximation for $\Delta _{r}$
depends on how well one can separate the Fourier components of
$i(k_{0})$. In the present case we applied this procedure only for
the sake of checking our results. An attempt to systematize this
method would be worth only if one's goals require an automated
phase-retrieval method. In our case, as we were interested in
giving a proof of principle only, the method of choice was not a
fully automated one, but a partially manual method which was
envisioned to demonstrate the feasibility of our approach.

\begin{figure}[tbp]
\begin{center}
\includegraphics[angle=0,scale=1.]{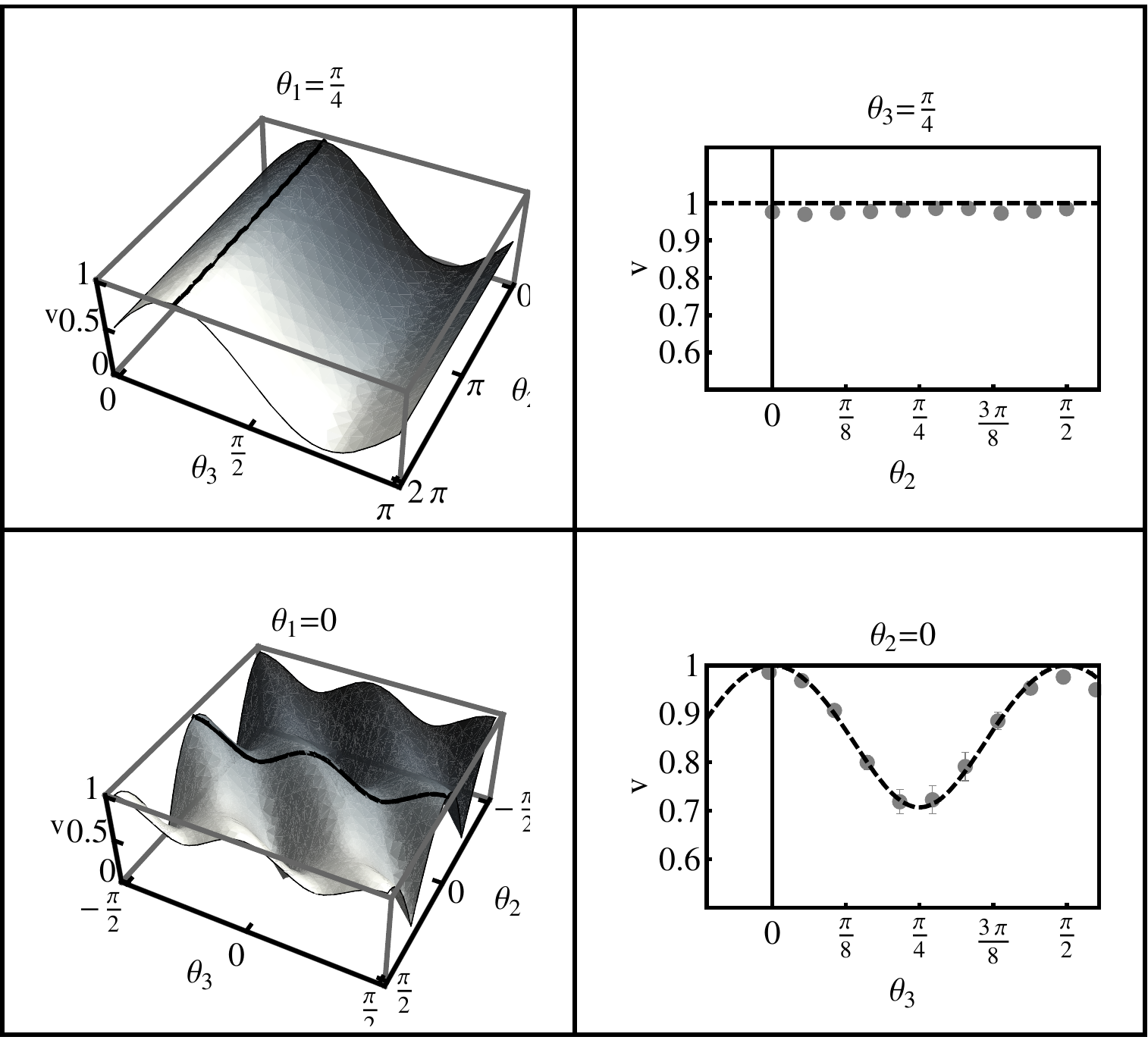}
\end{center}
\caption{Interferometric measurement of the visibility
$v(\theta_{1},\theta_{2},\theta_{3})$. The left panels show the
surfaces obtained by fixing one of the three angles, $\theta_{1}$,
as indicated. The right panels show the experimental results that
correspond to the curves highlighted on the surfaces. The upper
curve is obtained by fixing $\theta_{3}$ besides $\theta_{1}$, the
lower curve by fixing $\theta_{2}$ and $\theta_{1}$. In the upper
curve all experimental values fall below the predicted (maximal)
visibility of $1$. This is because $I_{min}$ is never zero, as
required to obtain $v=1$. By subtracting the nonzero average of
$I_{min}$ the experimental points would fall above and below the
theoretical curve, as it occurs for the lower curve, which
corresponds to $v<1$.} \label{vis}
\end{figure}

Another series of tests was devoted to measuring the visibility
$v$ as given in Eq. (\ref{vis2}). The quantity $v(\theta
_{1},\theta _{2},\theta _{3})$ was submitted to test by fixing two
of its three arguments. Our results are shown in Fig. (\ref{vis}).
The left panels
correspond to $v(\theta _{1},\theta _{2},\theta _{3})$ as a function of $%
\theta _{2}$ and $\theta _{3}$, that is, the surface obtained by fixing $%
\theta _{1}$ as indicated. In the right panels we compare the
theoretical predictions against our measurements of $v(\theta
_{1},\theta _{2},\theta _{3})$, whereby two of the three arguments
have been held fixed. The interferograms were evaluated following
a procedure similar to the one already explained. However, in this
case it was not the full cross section of the beam that was
submitted to evaluation, but a manually chosen region of the
images corresponding to a part of the input beam having almost
uniform intensity. This had to be so, because Eq. (\ref{vis2})
presupposes a uniform profile of the input beam. In order to test
the visibility of the whole cross section of the beam, Eq.
(\ref{vis2}) should be modulated with a Gaussian envelope. Such a
refinement was however unnecessary for our scopes. In any case,
the experimental value of the visibility, viz.,
$(I_{max}-I_{min})/(I_{max}+I_{min})$, was obtained by choosing in
each interferogram several maxima and minima, so as to assess the
accuracy of our measurements. Thus, the error bars in the figures
take proper account of the tiny variations in the chosen region of
the input-beam profile. As can be seen, the experimental values
closely fit the theoretical predictions.

\section{\label{sec:level8}Conclusions}

We have carried out theoretical calculations and the corresponding
measurements of Pancharatnam's phase by applying polarimetric and
interferometric methods. Our interferometric array is robust
against thermal and mechanical disturbances. It can be implemented
with a Michelson, a Sagnac or a Mach-Zehnder interferometer. We
have compared our measurements with those obtained in a
polarimetric array, finding similar results in both cases. Our
polarimetric array consisted of five wave-plates and two
polarizers. Five plates are necessary to realize an arbitrary
$SU(2)$ transformation with the polarimetric array. As well known,
three plates are instead required for realizing an arbitrary
$SU(2)$ transformation with an interferometric array. The whole
Poincar\'{e} sphere of polarization states could be explored with
both our polarimetric and interferometric arrays. Thus, any two
given polarization states could be connected by the appropriate
$SU(2)$ transformation. The associated relative Pancharatnam's
phase would thereby be realized. This phase can be decomposed as a
sum of a dynamical and a geometrical phase.  By appropriately
choosing the path connecting two given states on the Poincar\'{e}
sphere one can study different aspects of both the dynamical and
the geometrical phase.

We have also tested theoretical predictions concerning fringe
visibility when applying the interferometric method. Our
experimental findings were in very good agreement with theoretical
predictions. This is interesting not only on its own, but also in
view of extracting Pancharatnam's phase from visibility
measurements in the case of mixed states. Indeed, it has been
proved \cite{sjoqvist} that, for mixed states, fringe visibility
is a simple function of Pancharatnam's phase.

\section{Acknowledgments}

We wish to thank E. J. Galvez for his technical advise and for
kindly lending us some optical equipment. This work was partially
supported by DAI-PUCP (contract number DAI-2009-0010). R. W.
acknowledges financial support from the Deutsche
Forschungsgemeinschaft (WI 393/20-1 and WI 393/21-1).

\newpage

\newpage

\newpage

\newpage

\end{document}